\documentclass[prl,twocolumn,showpacs,preprintnumbers,amsmath,amssymb]{revtex4}
\usepackage{color,graphicx,epsfig}

\usepackage{natbib}
\usepackage{amsmath, amssymb, graphics, mathrsfs}

\newcommand{\mathsym}[1]{{}}

\usepackage{slashed}

\begin{document}
\title{Measurement of the $\nu=1/3$ fractional quantum Hall energy gap in suspended graphene}
\author{Fereshte Ghahari, Yue Zhao, Paul Cadden-Zimansky, Kirill Bolotin$^*$ and Philip Kim}
\affiliation{Department of Physics, Columbia University, New York, New York 10027}

\begin{abstract}
We report on magnetotransport measurements of multi-terminal suspended graphene devices. Fully developed integer quantum Hall states appear in magnetic fields as low as 2~T.  At higher fields the formation of longitudinal resistance minima and transverse resistance plateaus are seen corresponding to fractional quantum Hall states, most strongly for $\nu=1/3$.  By measuring the temperature dependence of these resistance minima, the energy gap for the 1/3 fractional state in graphene is determined to be at $\sim$20~K at 14~T.
\end{abstract}


\maketitle

In the low magnetic field regime, the integer quantum Hall (IQH) effect of graphene is marked by an anomalous half-integer quantum Hall conductivity $\sigma_{xy}=g_s(n+1/2){e}^{2}/h$, where $n$ is an integer and $g_s = 4$ is the Landau level (LL) degeneracy resulting from the degenerate spin and valley isospin degrees of freedom.  This anomalous quantum Hall conductivity led to the observation of the filling factor sequence $\nu=\pm2,\pm6,\pm10$~\cite{novo2,zhang1}. Subsequently, new broken-symmetry IQH states, corresponding to filling factors $\nu=0, \pm1, \pm4$ have been resolved in magnetic fields of $B>$ 20~T, indicating the lifting of the fourfold degeneracy of the LLs \cite{zhang2,zhang3}.  These filling factors have been suggested to be the result of various novel correlated states mediated by electron-electron (e-e) interactions~\cite{eeReviews}.

In the strong quantum limit, e-e interactions in 2-dimensional electron gasses (2DEGs) can lead to the fractional quantum Hall (FQH) effect~\cite{tsu}, many-body correlated states where the Hall conductance quantization appears at fractional filling factors. In recent investigations of transport properties in two-terminal high-mobility suspended graphene devices~\cite{kiril3,du2}, a quantized conductance corresponding to the $\nu=1/3$ FQH state has been observed, suggesting the presence of strong e-e interactions in this system. However, due to the inherent mixing between longitudinal and transverse resistivities in this two-terminal measurement~\cite{ab}, quantitative characterization of the observed FQH states such as the FQH energy gap is only possible in an indirect way~\cite{ab2}. Although multi-terminal measurements on suspended graphene samples have been reported previously~\cite{kiril1, du1}, the mechanical ~\cite{gu} or thermal instability ~\cite{ska} of these samples has precluded even the observation of a fully-quantized IQH effect.

\begin{figure}[b]
\centering
\includegraphics[width=1.0\linewidth]{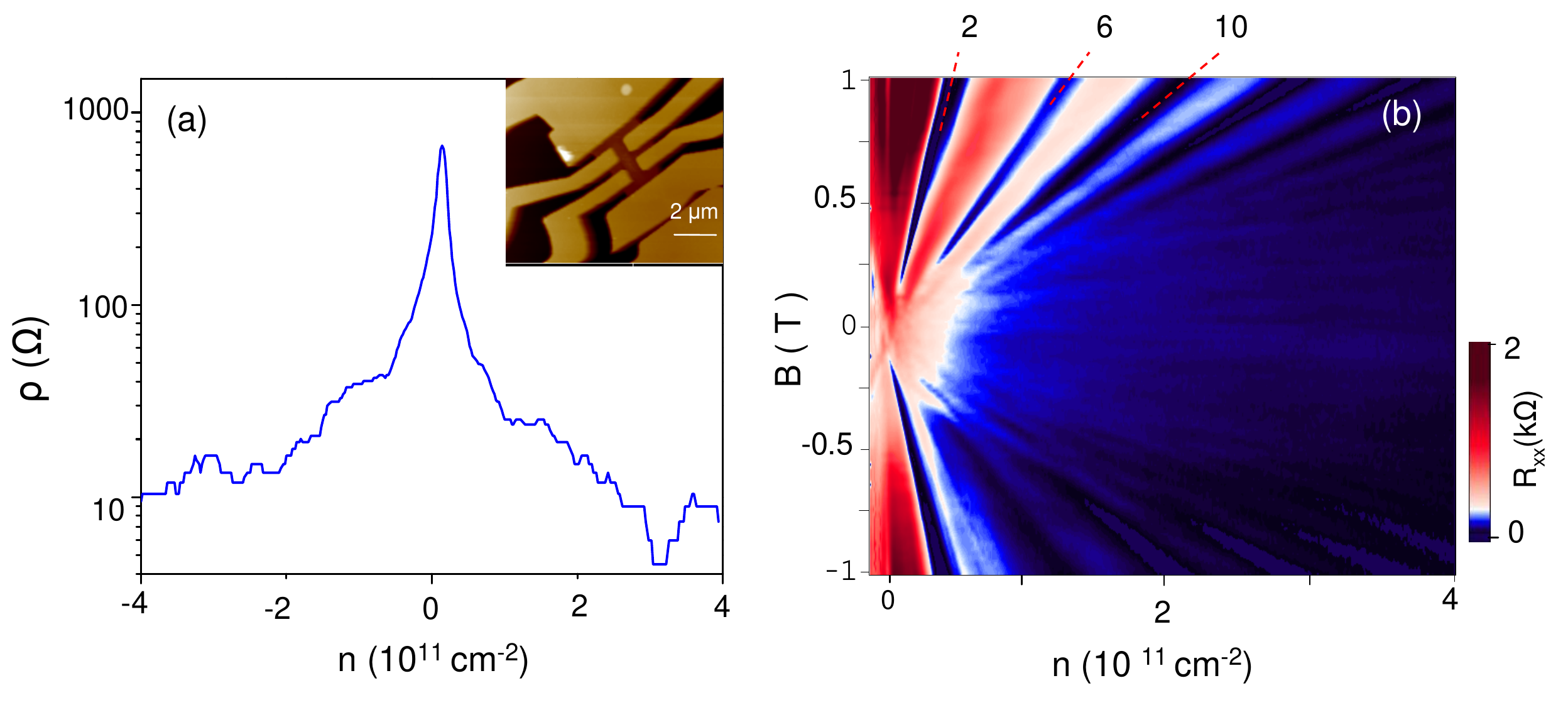}
\caption{(color online).(a) Resistivity of a suspended graphene sample (S1) versus carrier density induced by back gate (resistivity is displayed on a log scale). Inset: Atomic force microscope image of a typical suspended device. (b) Landau fan diagram $R_{xx}(V_g,B)$ at T=7~K measured in sample S3. The dark blue lines indicate minima in $R_{xx}$. The dashed lines with integers indicate the corresponding filling factors.}
\end{figure}

Recently, the improvement of graphene mobility up to 8~m$^2$/Vsec has been reported for substrate-supported graphene devices fabricated on a single-crystal hexagonal boron nitride substrates~\cite{Cory}. Multi-terminal transport measurements performed on such devices in magnetic fields up to 35~T reveal several FQH states whose filling factors are mostly integer multiples of $1/3$. The energy gaps of these states have been measured for $\nu>1$, exhibiting an unusual hierarchy among these FQH states~\cite{cory2}. However, the characterization of FQH states with $\nu<1$, notably the $1/3$ FQH state, could not be reliably conducted in these samples due to inhomogeneous broadening near the charge neutrality point. As stronger e-e correlations are expected for this lower density regime, further experiments on cleaner samples are desirable.

In this letter, we report on the measurement of multi-terminal IQH and FQH effects in ultraclean suspended graphene samples in this low density regime. Filling factors corresponding to fully developed IQH states, including the $\nu=\pm1$ broken-symmetry states and the $\nu=1/3$ FQH state are observed. The energy gap of the 1/3 FQH, measured by its temperature-dependent activation, is found to be much larger than the corresponding state found in the 2DEGs of high-quality GaAs heterostructures.

\begin{figure}[t]
\centering
\includegraphics[width=1.0\linewidth]{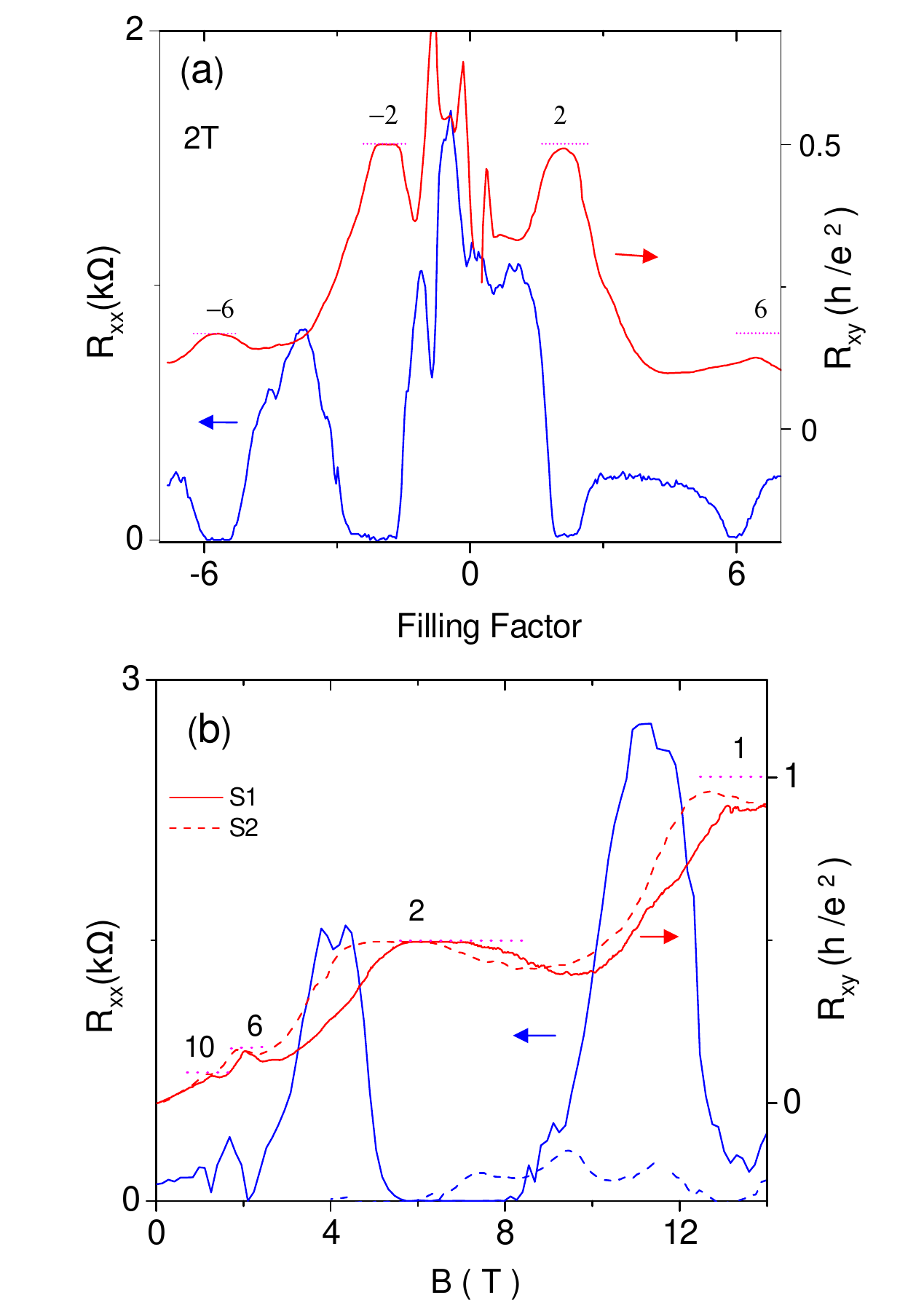}
\caption{(color online).(a) $R_{xx}$ and $R_{xy}$ measured as a function of filling factor $\nu$ at $B=$2~T and $T=1.7$~K for sample S1. -$R_{xy}$ is plotted for $\nu>0$.(b) $R_{xx}$ and $R_{xy}$ measured as a function of magnetic field $B$ at the hole density of $n=3.2\times 10^{11}$cm$^{-2}$ in samples S1 (solid line) and S2 (dashed line). Numbers above horizontal lines indicate the corresponding filling factors.}
\end{figure}
Our suspended graphene devices are fabricated using the method described in references~\cite{kiril3, kiril1}:  Mechanically exfoliated graphene samples are deposited on SiO$_2$/Si substrates. Electrical contacts are made using electron-beam lithography followed by the thermal evaporation of Cr/Au electrodes. The SiO$_2$ under the flake is subsequently removed via a chemical etch in buffered hydrofluoric acid. A typical multi-terminal graphene device with a lateral size $\sim$3~$\mu$m, suspended 150-200~nm above the SiO$_2$/Si substrate, is shown in the inset of Fig.~1(a).

We studied a total of three 4-terminal and one 6-terminal devices where better quality results were obtained in three of the samples, labeled S1, S2 and S3. The longitudinal resistance $R_{xx}$ and transverse Hall resistance $R_{xy}$ were measured as a function of the applied gate voltage $V_g$ on the doped Si back-gate, which tunes the carrier density $n$ in the graphene. The carrier density is determined using Hall measurements and $n(V_g) = B/{eR_{xy}(V_g,B)}$, where $B$ is the applied magnetic field. The gate voltage is limited to less than 10~V to avoid electrostatic collapse of the suspended devices. The initial mobility of as-fabricated devices is typically less than 1.5~m$^{2}$/Vs,  comparable to samples on substrate. Sending a large current density ($\sim$0.5~mA/$\mu$m) through a device heats up the graphene samples up to $\sim 400^{\circ}$ C~\cite{Berciaud}, which typically removes many of  adsorbed impurities, resulting in an extremely sharp peak in resistivity $\rho$ as $n$ changes from electrons to holes (Fig.~1(a)). In our study, the samples exhibit mobilities ranging from 8-15 m$^{2}$/Vs at the temperature $T=$ 1.7~K, where most of our data was taken. In these ultraclean suspended samples, the Shubnikov de Haas (SdH) oscillations, resulting from the quantized cyclotron orbits are observable at relatively low magnetic fields. Fig.~1(b) displays a Landau fan diagram where $R_{xx}$ is plotted as a function of $n$ and $B$ in the low magnetic field regime ($|B|<$ 1~T). In this diagram the SdH oscillation minima (later developing to the quantum Hall minima) appear as strips fanning out from the origin point $B=0$, $n=0$ with the slope $dn/dB=\nu{e}/h$. These strips survive in fields down to 0.1~T in our samples, in accordance with the mobility of $\sim$10 m$^{2}$/Vs calculated from conductivity measurements.

As $B$ increases, the observed SdH oscillations fully develop into the IQH effect. Fig.~2(a) shows $R_{xx}$ and $R_{xy}$ of device S1 as a function of filling factor $\nu=ne/hB$ (tuned by $V_g$) at a fixed magnetic field $B =$2~T.  A series of quantum Hall states,  i.e., zeroes in $R_{xx}$ and plateaus in $R_{xy}$, quantized to values $h /{\nu e^{2}}$ with integer filling factors $\nu=\pm 2, \pm 6$, are observed within the gate bias window.  More IQH statescan be observed using a field sweep at fixed gate voltage. Fig.~2(b) shows $R_{xx}$ and $R_{xy}$ measured as a function of magnetic field at a fixed hole density of $3.2 \times 10^{11}$~cm$^{-2}$ for S1 and S2. At least two well-defined plateaus with values $(h/6e ^{2})$ and $(h/2e ^{2})$ are observed,  while the $\nu =1 $ broken-symmetry IQH state is being reached $(h/e ^{2})$ at 14~T.  We note that the development of this $R_{xy}$ Hall plateau is not complete, measuring only $\sim$95 \% of the full quantization value. This deviation of the Hall resistivity from the expected quantization value may be attributable to the presence of non-ideal disordered contacts which can introduce a non-equilibrium population of edge states that perturbs the quantization for the small samples used here~\cite{but}.  It has also been suggested that the proximity between the current leads and the voltage probes could short-circuit the hall voltage in such small samples~\cite{ska}.

As we move to the low-density regime corresponding to $\nu < 1$, the FQH effect starts to be detected for  $B>$10~T in all three samples (S1, S2 and S3). Fig.~3(a) displays $R_{xx}$ and $R_{xy}$ plotted as a function of $B$ at a fixed density of $1.9 \times$10$^{11}$~cm$^{-2}$ (dotted line) and $0.96 \times$10$^{11}$~cm$^{-2}$ (solid line) respectively. At higher density, we notice that $R_{xy}$ increases further and $R_{xx}$ shows an additional dip at the field corresponding to $\nu \approx 2/3$. Upon further decreasing the density, lower filling fractions come into the observable window set by the maximum probing magnetic field (14~T), and $R_{xx}$ develops even deeper local minimum at the field corresponding to $\nu \approx 1/3$ with a plateau-like feature in $R_{xy}$. However, similar to the $\nu=1$ broken-symmetry IQH state, the corresponding features in $R_{xy}$ are not fully quantized in this low density and high Hall voltage regime. In sample S3 (Fig.~3(b)), in addition to minimum around $\nu \approx 1/3$, another minimum is visible around $\nu \approx 1/2$, but there is no feature close to a 2/3 filling fraction.

We note that, when scaled by the filling factor, these local minima of $R_{xx}$ are robust features at different magnetic fields and densities. Fig.~3(c) and Fig.~3(d) show the $R_{xx}$ of S1 and S2 as a function of filling fraction $\nu$ at different $B$ and $n$. The $R_{xx}$ traces exhibit local minima corresponding to filling fractions $\nu \sim $ 1/3, 1/2, 2/3 and 1. For all three samples, we observe strong minimum for $\nu=$ 1 and 1/3. But $\nu \sim$ 1/2 only shows up clearly for two samples S2 and S3 while $\nu \sim$ 2/3 emerges only in S1, indicating that these features are rather fragile and are more sample dependent compared to $\nu=$1/3. For example, in Fig.~3(d), in addition to the IQH state at $\nu=-1$ (feature C) two other notable features emerge as relatively deep minima in $R_{xx}$ (A and B), located at filling factors $\nu=-0.34$ and $\nu=-0.64$ respectively. Further confirmation on the nature of these states can be provided by means of a Landau fan diagram where $R_{xx}$ is plotted as a function of $V_g$ and $B$ (Fig.~3(e)). From the slopes of the lines in marked A and B we estimate that features A and B follow $\nu=0.31\pm 0.02$ and $\nu=0.66\pm 0.02$ lines, respectively. We thus assign features A and B to be the minima corresponding to the 1/3 and 2/3 FQH states.

\begin{figure}[t]
\centering
\includegraphics[width=1.0\linewidth]{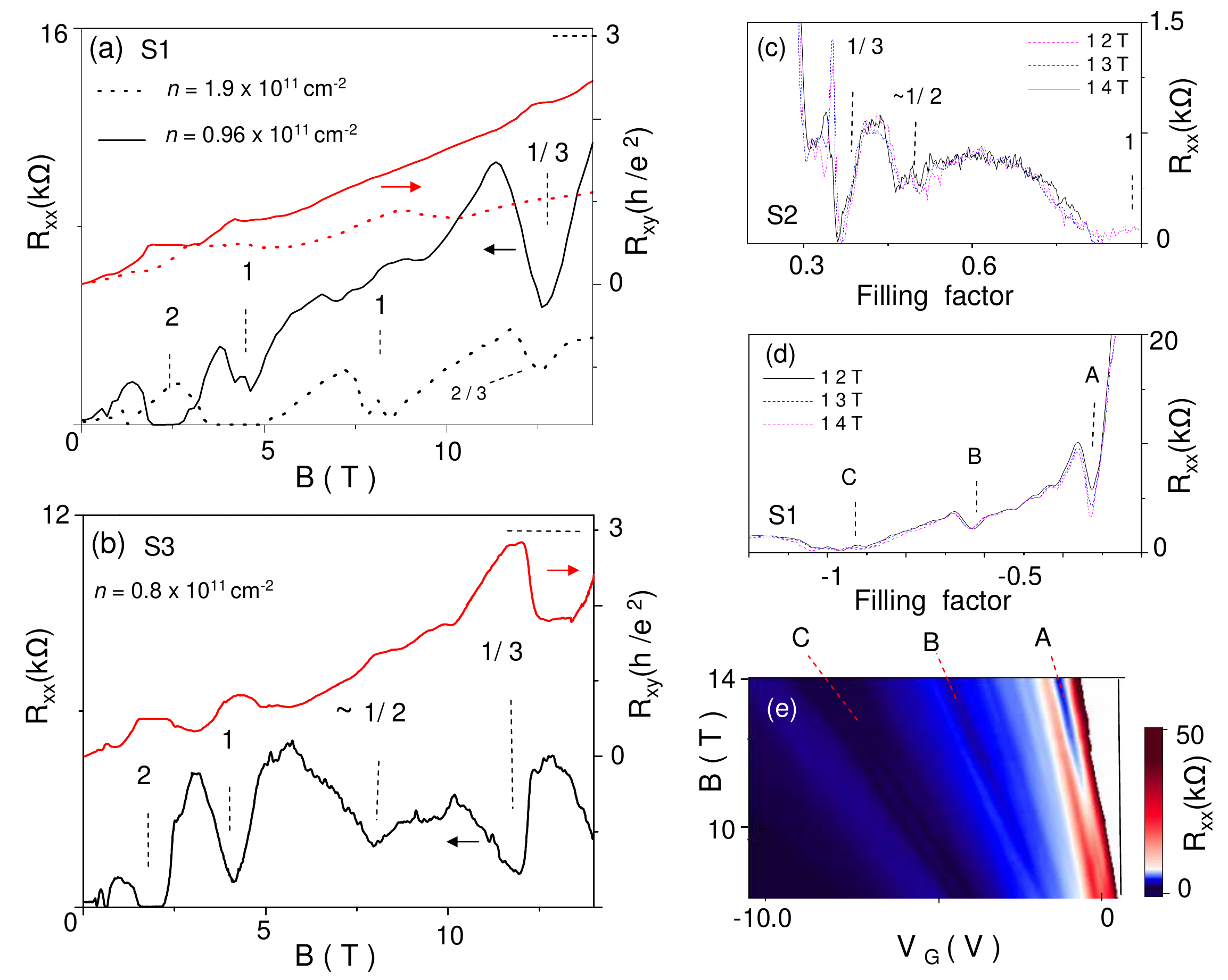}
\caption{(color online). (a) $R_{xx}$ and $R_{xy}$ measured in sample S1 as a function of $B$ at fixed density of $1.9 \times$10$^{11}$~cm$^{-2}$ (dotted line) and $0.96 \times$10$^{11}$~cm$^{-2}$ (solid line). (b)$R_{xx}$ and $R_{xy}$ measured in sample S3 as a function of magnetic field $B$ at fixed density of $0.8 \times$10$^{11}$~cm$^{-2}$ . Numbers with vertical lines indicate the corresponding filling factors. The straight horizontal lines indicates the expected $R_{xy}$ value for the 1/3 FQH state. (c) $R_{xx}$ as a function of filling factor at $B=12-14$~T measured in sample S2.(d)  $R_{xx}$ as a function of filling factor at $B=12-14$~T measured in sample S1.  The letters A, B and C with vertical lines indicate minima corresponding to filling factors 1, 2/3 and 1/3 respectively. (e) Landau fan diagram $R_{xx}(V_g,B)$ at T=1.7~K measured in sample S1. The dark blue lines indicate minima in $R_{xx}$. }
\end{figure}

The strongly developed minima in $R_{xx}$ for the 1/3 state now allows us to probe the energy gap associated with it. To quantify the energy of this FQH state, we measure the temperature dependence of $R_{xx}$.  Fig.~4(a) and (c) display $R_{xx}(V_g)$ measured at a sequence of different temperatures at 14~T for S1 and S2, respectively. The minimum corresponding to the 1/3 FQH state, $R_{xx}^{min}$, increases as $T$ increases. An Arrhenius plot for the $R_{xx}^{min}$ (Fig.~4(b)) shows an activation behavior indicated by $\log(R_{xx}^{min})\sim T^{-1}$ in the temperature range between 9-22~K for S1. At lower temperatures (2-8~K) $\log(R_{xx}^{min})$ deviates from a simple activation behavior , turning into the slower temperature dependence expected for variable-range hopping behavior. Similar trends were observed in early experiments on GaAs 2DEGs~\cite{boe}.

From the high-temperature activation behavior, the transport gap $\Delta E$ of the 1/3 FQH state can be obtained using $R_{xx}^{min}\sim \exp(-\Delta E/2k_BT)$ at a fixed magnetic field.  Fig.~4(d) shows $\Delta E$ as a function of $B$ determined from the line fits in Fig.~4(b) for sample S1.  A similarly determined activation energy gap for the 1/3 FQH state of sample S2 is also included in this figure (For this sample the $R_{xx}^{min}$ shows an activation behavior in the temperature range between 4-13~K). For both samples, we obtained similar magnitude of the energy gap, increasing with increasing $B$, as is expected.

Naively, the magnetic field dependence of energy gap of 1/3 FQH state would be expected to increase with $\sim\sqrt{B}$, considering the e-e interaction energy scales with the Coulomb energy scale $ e^{2}/{\epsilon l_B}$, where $\epsilon$ is the dielectric constant and $l_B$ is the magnetic length proportional to $1/\sqrt{B}$. As an alternative scenario, however, the activation energy gap could be linear in $B$, if the nature of charged quasiparticle excitations of this FQH state are associated with spin-flips in skyrmion-like excitations of a spin-polarized FQH ground state~\cite{deth}. Given the field range and error bars of the data points, we cannot rule out either scenario at present. We remark the linear fit (dashed lines) yields a negative y-intercept $\sim$ 20~K, which can be interpreted as the broadening $\Gamma$ of the fractional state.This value is slightly larger than $\hbar/ \tau \sim$ 4~K, the expected LL broadening estimated from the scattering time $\tau$ obtained from the mobility of the sample. However, the $\sqrt{B}$-fit (dotted lines) yields a negative y-intercept $\sim$ 60~K too large to be considered a reasonable $\Gamma$.

\begin{figure}[t]
\centering
\includegraphics[width=1.0\linewidth]{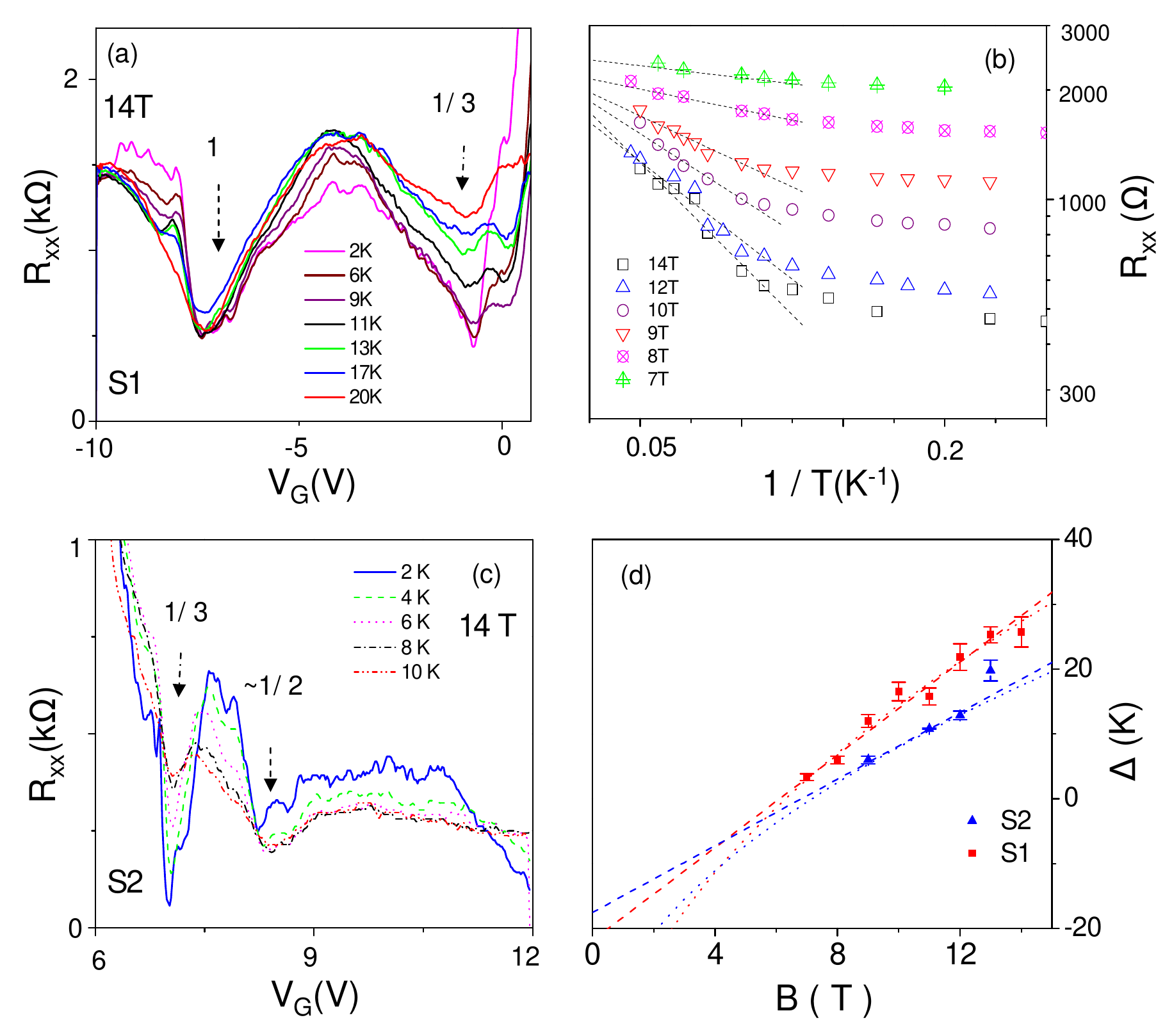}
\caption{(color online) (a) $R_{xx}$ measured at different temperatures at 14~T for sample S1. (b) Arrhenius plots of $R_{xx}^{min}$ versus $1/T$ at different fields at $\nu=1/3$ FQH state for sample S1.  The straight lines are linear fits to the data at the high-temperature activation range. (c) $R_{xx}$ measured at different temperatures at 14~T for sample S2. (d)  The activation gap as a function of field for two samples:  S1 and S2. The dashed lines are fits to $B$ and the dotted lines are fits to $\sqrt{B}$.}
\end{figure}

Considering the LL broadening $\Gamma \sim$20~K obtained from the linear fit, the intrinsic energy gap for $\nu=$1/3 is estimated to be $\Delta_{1/3}=\Delta E+\Gamma \approx$ 40-45~K at $B=$14~T. It has been predicted that the 1/3 state is both spin and valley-isospin polarized in the SU(4) configuration space~\cite{apal}, and calculations for this fully polarized 1/3 state have given a gap value of ${C~e^{2}}/{\epsilon l_B}$, with the numerical constant $C\approx$0.05-0.1 ~\cite{apal,toke1}. For graphene in vacuum, the dielectric constant from in-plane dynamic screening is estimated to be $\epsilon=$5.24~\cite{gon}. The predicted gap is then in the range $\Delta_{1/3}=$26-50~K at $B=$14~T, in reasonable agreement with the gap measured in our experiment. For comparison, the observed disorder reduced $\Delta_{1/3}$ in graphene is at least 3 times larger than that of the 2DEGs in the best quality GaAs heterojunctions in a similar field range~\cite{deth}. We further remark that $\Delta_{1/3}$ obtained in this experiment is much larger than the gaps obtained for FQH states associated to $\nu>1$ considering those gaps are obtained at $B=$35~T~\cite{cory2}. This comparison thus suggests an unusual robustness of the 1/3 state in graphene, inviting further investigation to elucidate its microscopic nature~\cite{Goerbig}.

In conclusion, we have measured the quantum Hall effect in multi-terminal suspended graphene devices. Both the broken symmetry IQH and 1/3 FQH states emerge, as manifested by minima in $R_{xx}$ and nearly quantized plateaus in $R_{xy}$.  From activation behavior of the $R_{xx}$ minima, the energy gap associated with the 1/3 FQH state is measured.

We thank J.~K. Jain for helpful discussions. This work is supported by the DOE (DE-FG02-05ER46215).

$^*$ Present address: Department of Physics, Vanderbilt University, Nashville TN 37212.

\end{document}